\documentclass[preprint,showpacs,preprintnumbers,amsmath,amssymb]{revtex4}
\usepackage{graphicx}
\usepackage{epsf}
\usepackage{dcolumn}
\begin{document}
\preprint{preprint - vortex group, iitb/tifr}
\title{Magnetization hysteresis and time decay measurements in FeSe$_{0.5}$Te$_{0.5}$ : Evidence for fluctuation in mean free path induced pinning}
\author{P. Das$^{1, 2, 3}$, Ajay D. Thakur$^1$,\footnote{Corresponding Author \\ Email: ajay@phy.iitb.ac.in; adthakur@gmail.com}, Anil K. Yadav$^1$, C. V. Tomy$^1$, M.R. Lees$^4$, G. Balakrishnan$^4$, S. Ramakrishnan$^2$, A. K. Grover$^2$}
\affiliation{$^1$ Department of Physics, Indian Institute of Technology Bombay, Mumbai 400076, India\\
$^2$ DCMP\&MS, Tata Institute of Fundamental Research, Homi Bhabha Road, Colaba, Mumbai 400005, India\\
$^3$ Institute of Materials Science, The University of Tsukuba, 1-1-1, Tennodai, Ibaraki 305-8573, Japan\\
$^4$ Department of Physics, University of Warwick, Coventry CV4 7AL, UK}
\date{\today}
\begin{abstract}
We present results of magnetic measurements relating to vortex phase diagram in a single crystal of FeSe$_{0.5}$Te$_{0.5}$ which displays second magnetization peak anomaly for $H \parallel c$. The possible role of the crystalline anisotropy on vortex pinning is explored via magnetic torque magnetometry. We present evidence in favor of pinning related to spatial variations of the charge carrier mean free path leading to small bundle vortex pinning by randomly distributed (weak) pinning centers for both $H \parallel c$ and $H \perp c$. This is further corroborated using magnetization data for $H \parallel c$ in a single crystal of FeSe$_{0.35}$Te$_{0.65}$. Dynamical response across second magnetization peak (SMP) anomaly in FeSe$_{0.5}$Te$_{0.5}$ has been compared with that across the well researched phenomenon of peak effect (PE) in a single crystal of CeRu$_2$.
\end{abstract}
\maketitle
The discovery of superconductivity in quaternary (1111) Iron (Fe) pnictide system LaFeAsOF at 26\,K \cite{kamihara} opened the flood gates for explorations on Fe based superconducting systems. Vigorous research in related systems has already demonstrated the existence of superconductivity in several Fe-based compounds including the ThCr$_2$Si$_2$-structure based quaternary (Ba,Sr)$_{1-x}$K$_x$Fe$_2$As$_2$ (122) \cite{rotter}, the ternary LiFeAs (111) \cite{tapp} and the binary FeSe/Te (11) \cite{wu}. In an uncanny resemblance to the importance of Cu-O planes in the cuprate high $T_{\rm c}$ superconductors, the FeAs or, FeSe/Te layers play a vital role in Fe-based superconductors. Amongst them the tetragonal FeSe/Te system has proved to be promising to understand the basic mechanism of superconductivity in Fe-based materials. The Fermi surface of the tetragonal system is very similar to that reported for the FeAs based superconductors \cite{fermi}, comprising cylindrical electron sections at the zone corners, cylindrical hole surface sections, and small hole sections at the zone center. Furthermore, these surfaces are separated by a $2D$ nesting vector at ($\pi$, $\pi$), another characteristic reminiscent of the FeAs based superconductors. Despite an apparent structural simplicity of FeSe/Te vis a vis other Fe-based systems, its physics is already shown to be both rich as well as interestingly complex. There have been reports hinting towards the possibility of an anisotropy in the symmetry of order parameter \cite{order}, multiple band gaps \cite{gap} and a dominant Pauli paramagnetic effect in the upper critical field(s) ($H_{c2}$) \cite{khim}. Also, the high pressure studies have shown an enhancement in $T_{\rm c}$ to as high as 36\,K in the FeSe system at a pressure of 38\,GPa \cite{press1, press2}. Using muon-spin-spectroscopy studies, the temperature dependence of the penetration depth in FeTe$_{0.5}$Se$_{0.5}$ was seen to be compatible with either a two-gap $s+s$-wave or an anisotropic s-wave model \cite{gb}. In view of the promising fabrication of superconducting wires of FeSe/Te by powder-in-tube technique \cite{nims}, it becomes important to understand its vortex phase diagram and the pinning mechanism. Prozorov et al. \cite{proz1} have reported the dynamical response of the flux line lattice via isothermal $M$-$H$ scans and magnetic relaxation measurements in Ba(Fe$_{0.93}$Co$_{0.07}$)$_2$As$_2$ and found a crossover from the collective to plastic creep regime near the peak position of the fishtail feature. Yadav et al \cite{yadav1, yadav2} have reported magnetic and transport studies in FeTe$_{0.6}$Se$_{0.4}$ which has an optimal $T_{\rm c}$ in the FeSe/Te system. It is indeed of interest to explore vortex physics in samples with a smaller $T_{\rm c}$ (away from optimal composition). Discovery of still higher $T_{\rm c} \sim 30$\,K in K$_{0.8}$Fe$_2$Se$_2$ \cite{guo} and proposals for enhancing $T_{\rm c}$ via suitable substitution in excess at the Fe site \cite{anil} have made vortex state studies on FeSe based systems all the more desirable. Magnetic relaxation studies are warranted to explore the pinning mechanism as well as its possible connection to crystalline anisotropy. Relaxation studies have been performed in the past on a host of low-$T_{\rm c}$ and high-$T_{\rm c}$ superconducting materials. In this paper we report detailed magnetization measurements on a single crystal of FeSe$_{0.5}$Te$_{0.5}$ with a $T_{\rm c} = 14.3$\,K. The results include: (i) observation of the second magnetization peak (SMP) (i.e., a fishtail feature), (ii) calculation of crystalline anisotropy based on torque magnetometry measurements, (iii) the estimation of flux pinning force density ($F_{\rm p}$), (iv) obtaining the vortex phase diagram, and (v) magnetic relaxation across SMP. Based on the above results, we try to provide an understanding of the underlying pinning mechanism and compare and contrast the results of relaxation studies with our earlier results in a weakly pinned crystal of CeRu$_2$ \cite{daeceru2}, which displayed peak effect phenomenon.\\

The FeSe$_{x}$Te$_{1-x}$ ($x=0.5$ and $x=0.35$) single crystals used for the present work were grown using the modified Bridgman method. The sensitivity to growth conditions must be kept in mind \cite{tsurkan}. In our case, a stoichiometric mixture of Fe powder (99.999\%), Se shots (99.99\%) and Te powder (99.999 \%) was sealed in an evacuated quartz tube (10$^{-6}$ mbar) and heated at a rate of 60$^o$C/h to 650$^o$C, kept for 48 h and then furnace cooled to room temperature. The sample was then homogenized and resealed in a quartz tube tapered at one end, heated at a rate of 60$^o$C/h to 970$^o$C, kept for 24 h and then cooled to 300$^o$C at 2$^o$C/h. The furnace is then cooled to room temperature. During the second heating, the quartz tube was kept in another quartz tube at high vacuum. Powder x-ray diffraction was performed at various stages of sample preparation using Philips powder x-ray diffractometer. For a majority of magnetization measurements reported in this paper, we chose a parallelepiped shaped piece of FeSe$_{0.5}$Te$_{0.5}$ weighing 22.5\,mg and having a superconducting transition temperature ($T_{\rm c}$) of 14.3\,K. In order to validate the key results propounded in this paper, we performed a set of measurements in another crystal of FeSe$_{0.35}$Te$_{0.65}$ with a $T_{\rm c}$ of 11.5\,K. The DC magnetization studies were performed using PPMS-VSM and SQUID-VSM, Quantum Design Inc., (USA) for field applied both parallel and perpendicular to the $c$-axis. The scan amplitude was chosen to be 2.0\,mm and 1.0\,mm for the PPMS-VSM and SQUID-VSM, respectively. Torque magnetometry measurements were performed on the sample using a suitably calibrated torque lever chip on a PPMS Tq-MAG (Quantum Design Inc., USA) option.\\

The single crystals of FeSe$_{x}$Te$_{1-x}$ ($x=0.5~{\rm and}~0.35$) used in the present study have layered planes held together by weak van der Waals interaction and can thus be cleaved easily. X-ray diffraction pattern obtained for one such cleaved single crystal piece is shown in Fig.~1. Prominent (00l) reflections are observed indicating that the c-axis of the single crystal is perpendicular to the cleaved surface. 
The panel (a) of Fig.~2 shows typical five quadrant isothermal magnetization hysteresis ($M$-$H$) loops recorded at several temperatures between 5\,K and 14\,K and for $H\parallel {\rm c}$. The minimum in magnetization located at a field value little above nominal zero field in a given $M$-$H$ loop represents the first magnetization peak characteristic, which amounts to (near) full penetration of the applied field in the bulk of the sample after zero field cooling. Thereafter, a prominent SMP, also known as the fishtail effect (FE), can be observed in the $M$-$H$ data at 7\,K, 9\,K and 11\,K. The onset ($H_{\rm SMP}^{\rm on}$ ) and peak ($H_{\rm SMP}^{\rm p}$) positions of SMP are marked for the 5\,K plot. Both $H_{\rm SMP}^{\rm on}$ and $H_{\rm SMP}^{\rm p}$ are seen to decrease along with the hysteresis width as the temperature enhances from 5\,K towards 12\,K. At 14\,K, $M$-$H$ loop, on repeated cycling, has a shape akin to that in a magnetically ordered system. However, the onset of diamagnetic response, pertaining to the onset of superconductivity can be identified by using the deviation of linearity criterion (from the paramagnetic normal state response), while decreasing the field from above the upper critical field ($H_{c2}$) \cite{shikha}. We could determine $H_{c1}$ values from the virgin portion (after zero field cooling) of the $M$-$H$ curves (up to 12.5\,K), using the deviation of linearity criterion from the Meissner response (data not shown). In panel (b) of Fig.~2, we present the $M$-$H$ loops recorded at several temperatures for $H\perp {\rm c}$. The scenario here is quite different when compared to the situation for $H\parallel {\rm c}$. Here, we do not observe a clear signature of SMP within our measurement range of field values. This essentially suggests an anisotropy in the pinning behavior of the flux line lattice across SMP.
In view of distinct differences in the observed scenario in the cases for field parallel and perpendicular to the c-axis, respectively from the perspective of the observation of SMP, we were led to the issue of the possible role of crystalline anisotropy and its effect on the vortex lattice structures. A direct way to experimentally obtain the anisotropy parameter $\gamma$ ($=\frac{H_{c2}^{ab}}{H_{c2}^{c}}=\frac{\xi_{ab}}{\xi_c}=\frac{\lambda_c}{\lambda_{ab}}=\frac{m_c^{\ast}}{m_{ab}^{\ast}}$ for a conventional, single-band, $s$-wave superconductor) is via measuring the torque acting on a superconducting sample in the mixed state with the magnetic field being applied at various angles with respect to the crystalline axis. We performed torque magnetometry measurements on our single crystal of FeSe$_{0.5}$Te$_{0.5}$. Figure 3 shows the torque ($\tau(\theta)$) data obtained at a temperature of 11\,K at 5\,kOe. In the field range $H_{c1} \ll H \ll H_{c2}$, the torque density is given by, $\tau = \frac{\phi_0B(\gamma^2-1)\sin 2\theta}{64\pi^2\lambda^2\gamma^{1/3}\varepsilon(\theta)}\ln \frac{\eta H_{c2, a}}{\varepsilon(\theta)}$ \cite{kogan88, farrel}. Here, $\theta$ is the angle between the magnetic induction $B$ and the crystalline $c$-axis, $\varepsilon(\theta) = \sqrt{\sin^2\theta + \gamma^2 \cos^2\theta}$, $\lambda^3 = \lambda_a^2\lambda_c$ and $\eta \sim 1$. The torque data is fitted using the above equation yielding an anisotropy parameter of 3.17. The value obtained for $\gamma$ is consistent with the results reported by Bendele {\it et al} \cite{bendele} in a single crystal of FeSe$_{0.5}$Te$_{0.5}$.

We extract critical current density values from the isothermal $M$-$H$ data making use of the Bean's critical state model formalism \cite{bean}, where, $J_c = 20 \frac{\Delta M}{a(1- \frac{a}{3b})}$ \cite{poole}, with $a~(=1.8\,mm)$ and $b~(=3\,mm)$ being the sample dimensions perpendicular to the field direction, and $\Delta M$ is the difference between the magnetization measured during the return and the forward legs of the $M$-$H$ loop. The main panel in Fig.~4 shows the plot of the normalized critical current density $J_c^{\rm norm}(H)$ at the temperatures of 5\,K, 7\,K, 9\,K and 11\,K. Here, the normalization has been done with the $J_c$ value at the peak position of SMP (where the correlation volume of the vortex lattice is expected to attain a minimum value \cite{shikha}). In the case of the 9\,K data, $H_{\rm SMP}^{\rm on}$ and $H_{\rm SMP}^{\rm p}$ are marked by arrows. The inset in Fig.~4 shows a color scale contour plot of $J_c (H,t)$, where $t=T/T_{\rm c}$ is the reduced temperature.
In order to understand the nature of pinning in more detail, it is useful to look at the variation of pinning force density, $F_{\rm p}$ with the magnetic field. In Fig.~5, we show the plot of normalized pinning force density ($F_{\rm p}^{\rm norm}$) as a function of reduced magnetic fields $h$ ($=H/H_{\rm c2}$, where $H_{\rm c2}$ is the upper critical field) at 10\,K and 11\,K. Note that the $F_{\rm p}^{\rm norm}$ curves for the two temperature values collapse into a unified curve. We fit these data within the Dew-Hughes scenario ($F_{\rm p} \sim h^{\alpha}(1-h)^{\beta}$) \cite{dew1, dew2}. The Dew-Hughes fit is shown by the dark violet line in Fig.~5,  and it yields the following values of the exponents : (a) $\alpha = 1.65$, and (b) $\beta = 2.95$. Here, it should be noted that the ratio $\frac{\alpha}{\alpha + \beta} \approx 0.358$ agrees well with the observed value of $h_{max}$ in accordance with the Dew-Hughes analysis \cite{dew1, dew2}. In the case of a system dominated by point pinning alone, $\alpha = 1$ and $\beta = 2$ with $F_{\rm p}^{\rm max}$ occurring at $h_{\rm max} \approx 0.33$ \cite{dew1, dew2}. In contrast, the grain boundary pinning is expected to lead to $h_{\rm max} \approx 0.2$, whereas, pinning due to variations in the superconducting order parameter leads to $h_{\rm max} \approx 0.7$ \cite{dew1, dew2}. In our case, $h^{\rm max} \approx 0.36$, implying that point pins alone can not rationalize the observed scenario. In the case of BaFe$_{1.8}$Co$_{0.2}$As$_2$, $h_{\rm max} \approx 0.45$, suggesting a possible correlation with inhomogeneous distribution of Co ions \cite{yamamoto}. Sun {\it et al} \cite{sun} did a similar analysis for a number of electron doped and hole doped iron-arsenide based superconductors, which included Ba$_{0.68}$K$_{0.32}$Fe$_2$As$_2$ ($h_{\rm max} \approx 0.43$), BaFe$_{1.85}$Co$_{0.15}$As$_2$ ($h_{\rm max} \approx 0.37$) and BaFe$_{1.91}$Ni$_{0.09}$As$_2$ ($h_{\rm max} \approx 0.33$). They observed that $H_{c2}$ and $H_{\rm SMP}^{\rm p}$ decreased faster with decreasing temperatures for the Ni substituted sample than for the K/Co substituted samples. Their observation was corroborated by the larger $\Delta T_{\rm c}$ for the Ni-substituted sample. However, the K-substituted sample showed the strongest pinning amongst the three systems, suggesting that an inhomogeneous distribution of dopant can not by itself explain the strong pinning in iron-arsenide based superconductors. Magnetic decoration experiments by Vinnikov {\it et al} \cite{vinnikov} in a variety of iron arsenide superconductors showed a disordered vortex state, whose nature was seen to be independent of the crystal structure type, doping and synthesis methods. The absence of an Abrikosov vortex lattice upto field values of 200\,Oe in their experiments suggests the dominance of small bundle pinning at low fields in these classes of superconductors. The intrinsic mechanism of pinning in these materials however remains intriguing \cite{inosov, eskildsen}. Yadav {\it et al} \cite{yadav2} obtained a value of $h_{\rm max}=0.28$ for the single crystal of FeSe$_{0.4}$Te$_{0.6}$. The variation in the value of $h_{\rm max}$ could be a result of changes in pinning force arising due to relative changes in the $Se$ and $Te$ compositions. It should also be kept in mind that the Dew-Hughes analysis is valid for $h$ obtained via field normalizations with respect to the upper critical field $H_{\rm c2}$. Several reports used a field normalization with respect to the irreversibility field, $H_{\rm irr}$ \cite{yadav2, yamamoto, sun}. Here, $H_{\rm irr}$ is the field value where $J_c (H)$ is measurably zero and the magnetization response is reversible for fields $H > H_{\rm irr}$. The discrepancies arising due to a different criteria \cite{yadav2, yamamoto, sun} used for the estimation of $h$ must therefore be dealt with care (more so because $H_{\rm irr}$ and $H_{c2}$ could have different temperature dependences \footnote{We would like to acknowledge one of the anonymous referee of Physical Review B for pointing this out.}). The observation of $H_{\rm irr} (T)$ values lying well below the $H_{\rm c2} (T)$ values over the entire H-T vortex phase diagram has been a distinct feature seen in a wide variety of low-T$_{\rm c}$ and high-T$_{\rm c}$ superconductors \cite{sarkar, sarkar1, sarkar2, ban, luirsi, salem, prando}. Here it is worthwhile to look at its ramifications on the exponents within the Dew-Hughes scenario for the iron chalcogenide superconductor under consideration. In Fig. 6 (a), we present the variation of $F_{\rm p}^{\rm norm}$ as a function of reduced field, $h$ ($=H/H_{c2}$) within the Dew-Hughes formalism when $\alpha$ is fixed to the best fit value and $\beta$ is varied. A similar plot is shown in Fig. 6 (b) where, $\beta$ is fixed to the best fit value and $\alpha$ is varied. It should be noted that the magnitude of $F_{\rm p}^{\rm norm}$ is suppressed at the high field end ($h$ closer to 1) when either $\alpha$ is reduced (cf. Fig. 6 (b)) or $\beta$ is increased (cf. Fig. 6 (a)) about the best fit values for $\alpha$ and $\beta$. To illustrate this point more clearly, it is useful to conceive a quantity $h_{\rm \Delta}$ (defined as the reduced field value at which $F_{\rm p}^{\rm norm} (h_{\rm \Delta})$ attains a value of 0.05). Thus $h_{\rm \Delta}$ could be considered to mimic the reduced irreversibility field $h_{\rm irr}$ (defined here via an empirical lower cut-off on $F_{\rm p}^{\rm norm}$ and hence on $J_c$). The insets in panels (a) and (b) shows the variation of $h_{\rm \Delta}$ as the magnitudes of the exponents $\alpha$ and $\beta$ are varied about the best fit values. This suggests that the occurrence of an $H_{\rm irr}(T)$ line well below the $H_{\rm c2}(T)$ line indeed affects the exponents $\alpha$ and $\beta$ within the Dew-Hughes formalism.

Nature of pinning in type-II superconductors can be broadly classified into two categories: (i) the one arising because of the spatial fluctuations in the transition temperature, $T_{\rm c}$ (known as $\delta T_{c}$ pinning) across the sample, and (ii) the other caused by the spatial variations in the charge carrier mean free path, $l$ (known as $\delta l$ pinning). In the case of $\delta T_{\rm c}$ pinning, the normalized critical current density, $J_c(t)/J_c(0) = (1-t^2)^{7/6}(1+t^2)^{5/6}$, while for $\delta l$ pinning, $J_c(t)/J_c(0) = (1-t^2)^{5/2}(1+t^2)^{-1/2}$, where $t=T/T_{c}(0)$ \cite{blatter, prl1994}. In Fig.~7 (a), we first plot the normalized $J_c(t)$ data (for $H \perp {\rm c}$) at 0\,kOe (the so called remanent state shown by filled hexagons), 5\,kOe (open stars), 10\,kOe (open diamond) and 20\,kOe (open triangles) for FeSe$_{0.5}$Te$_{0.5}$. These have been extracted from the $J_c(H)$ plots at various reduced temperature values, $t$ and thereafter normalized using the $J_c(0)$ values obtained from the fit to the expression for $\delta l$ pinning. The theoretical estimates of $J_c(t)/J_c(0)$ within the scenarios of $\delta T_{\rm c}$ pinning and $\delta l$ pinning are shown by bold lines. The observations point to the dominance of the $\delta l$ pinning mechanism in FeSe$_{0.5}$Te$_{0.5}$, suggesting the occurrence of single vortex pinning by randomly distributed weak pinning centers. In Fig.7 (b), we present a similar analysis for $H \parallel c$, where again the $\delta l$ pinning mechanism is evident.  Here it should be noted that whether the signature of SMP is conspicuously present (for $H \parallel c$) or absent ($ H\perp c$) in isothermal M-H loops, the variation of $J_c(t)$ points to a predominant $\delta l$ pinning mechanism in FeSe$_{0.5}$Te$_{0.5}$. We performed a similar analysis for the pinning mechanism for the single crystal of FeSe$_{0.35}$Te$_{0.65}$, where the signature of SMP is less conspicuous in isothermal M-H loops over the entire range of measurement temperatures. The panel (a) of Fig.8 shows the isothermal M-H loops for FeSe$_{0.35}$Te$_{0.65}$ (for $H \parallel c$) at temperatures of 3\,K, 5\,K, 7\,K, 9\,K and 11\,K, respectively. In Fig. 8 (b) we plot the normalized $J_c(t)$ data at 0~kOe, 5~kOe and 10~kOe, respectively along with the theoretically expected fits for $\delta T_{\rm c}$ and $\delta l$ pinning scenarios. Here again the experimental data fit better to the curve corresponding to the $\delta l$ pinning picture. Our observation of a dominant $\delta l$ pinning in FeSe$_{0.5}$Te$_{0.5}$ is in contrast to a recent claim of a prevalent $\delta T_{\rm c}$ pinning in FeSe$_{0.40}$Te$_{0.60}$ by Liu {\it et al} \cite{epl2010}.  In their report, however, Liu {\it et al} \cite{epl2010} have not attempted to make a quantification of the temperature variation of normalized critical current density. They attribute the broadening of the SMP as a signature of $\delta T_{\rm c}$ pinning. One may note that the possible fingerprint(s) of analogue of SMP anomaly evident in isothermal plots (cf. Fig. 2 (a) and Fig. 4) is not present in the extracted isofield data (see Fig. 7 (b)) at $H=$ 5\,kOe and 10\,kOe (for $H \parallel c$). Similar apparent difficulty was encountered in reconciling M-H data showing fishtail anomaly in high $T_{\rm c}$ samples with the extracted isofield plots for field values lower than the onset field of SMP anomaly. 

Vortex phase diagrams have been presented based on magnetization and transport studies in a variety of Fe based superconductors including Ba$_{1-x}$K$_x$Fe$_2$As$_2$ \cite{yang, salem}, Ba(Fe$_{0.93}$Co$_{0.07}$)$_2$As$_2$ \cite{proz1}, SmFeAsO$_{0.8}$F$_{0.2}$ \cite{prando}, FeSe$_{0.40}$Te$_{0.60}$ \cite{yadav2}, etc. In a variety of low T$_{\rm c}$ and high T$_{\rm c}$ superconductors, the vortex phase diagram is constructed based on the characteristic fields obtained from anomalous variations of J$_{\rm c}$ \cite{shikha, sarkar, adtprb, adtjpsj}. In Fig.~9, we plot such a field-temperature phase diagram for the case of FeSe$_{0.5}$Te$_{0.5}$ with $H\parallel {\rm c}$. The locations of $H_{\rm SMP}^{\rm on}$, $H_{\rm SMP}^{\rm p}$, $H_{\rm irr}$ and $H_{c2}$ (as obtained via isothermal $M$-$H$ measurements) are shown by open square, open circle, filled circle and filled diamond symbols, respectively. Following Prozorov {\it et al} \cite{proz1}, we attempted to make a least square fit to the expression $H_x(T) = H_x(0)[1-(\frac{T}{T_{\rm c}})^p]^n$. The least square fits yield the following results for the various characteristic fields : (i) for the onset of SMP, $p=1$, $n=\frac{4}{3}$ and $H_{\rm SMP}^{\rm on}(0) = 17.45~kOe$, (ii) for the peak of SMP, $p=1$, $n=\frac{3}{2}$ and $H_{\rm SMP}^{\rm p}(0) = 123.8~kOe$, (iii) for the irreversibility line, $p=2$, $n=\frac{9}{5}$ and $H_{\rm irr}(0) = 220~kOe$, and (iv) for the upper critical field line, $p=1$, $n=\frac{4}{3}$ and $H_{\rm SMP}^{\rm on}(0) = 400~kOe$. The coefficients $p$ and $n$ are identical to the results of Prozorov {\it et al} \cite{proz1} for $H_{\rm SMP}^{\rm p}$ and $H_{c2}$, however, for $H_{\rm irr}$, $n=\frac{9}{5}$ (unlike the value of $\frac{3}{2}$ obtained for Ba(Fe$_{0.93}$Co$_{0.07}$)$_2$As$_2$). Similar power-law behavior for the onset and peak fields for SMP were also observed in the case of YBCO \cite{klein} and Ba$_{0.6}$K$_{0.4}$Fe$_2$As$_2$ \cite{ yang}. It should, be noted that the temperature dependences of the $H_{\rm SMP}^{\rm on}$ and $H_{\rm SMP}^{\rm p}$ lines (see Fig.~9) are similar to those reported for YBCO \cite{klein}, whose pinning properties were seen to be dominated by the $\delta l$ pinning mechanism \cite{prl1994}. 

In view of the absence of any report on observation of an ordered vortex lattice in Fe-based superconductors via small angle neutron scattering, we refrain from putting any labels on the various phases in the vortex phase diagram. Here, exploring vortex dynamics in FeSe based superconductors has the potential of shedding some additional light on the pinning behavior and occurrence of order-disorder transformation in its vortex matter. Peak effect (PE) and SMP are well researched attributes associated with phase transformations in vortex matter. While the former is seen to occur close to the $H_{c2}(T)$ line in the vortex matter phase diagram, the latter is observed deep within the mixed state \cite{sarkar}. Both are associated with a concominant increase in the vortex pinning energy and hence an anomalous modulation in $J_c$. Changes in dynamical time scales across these characteristics are anticipated and are seen to occur in a wide variety of experiments in low $T_{\rm c}$ and high $T_{\rm c}$ superconductors \cite{yeshurun, maple, kalisky, adtphysica}. Taking cue from such observations, we attempt to make a similar analysis across the SMP in our single crystal of FeSe$_{0.5}$Te$_{0.5}$ and compare it with that in a single crystal of another low $T_{\rm c}$ superconductor CeRu$_2$ ($T_{\rm c} \approx 8$\,K), which displays only the PE phenomenon \cite{maple}. In particular, it is useful to look at the normalized magnetic relaxation rate, $S$, given by $S=\mid\frac{d \ln M}{d \ln t}\mid$ \cite{kalisky}. In the case of CeRu$_2$, Tulapurkar {\it et al} \cite{ashwin} demonstrated the occurrence of a jump in equilibrium magnetization illustrating the presence of a first-order transformation in the vortex lattice across PE. Also, within the collective pinning scenario, there is a large change in the correlation volume ($V_{\rm c}$) across the PE \cite{lo}. It is of interest to compare temporal decay response of magnetization across SMP in FeSe$_{0.5}$Te$_{0.5}$ with that across PE in CeRu$_2$ within the light of the results of Kalisky {\it et al} \cite{kalisky} and Thakur {\it et al} \cite{adtphysica}. Panel (a) of Fig.~10 shows a portion of the two quadrant isothermal $M$-$H$ loop at 9\,K for FeSe$_{0.5}$Te$_{0.5}$ single crystal. The inset in Fig.~10(a) shows the time decay ($M(t)$) data at 27\,kOe and 9\,K (measured up to 10$^4$\,s). The magnetization values at 10$^4$\,s for various field values along the return leg are shown by open red symbols (with dotted red line as a guide to eye for the portion of the time evolved $M$-$H$ loop at 10$^4$\,s). Figure 10(b) shows a plot of $S(H)$ at 9\,K across the SMP for the same sample. Modulation in $S$ appears to have little correlation with the modulation of $M$ across SMP. Figure 11(a) shows a portion of the $M$-$H$ loop for CeRu$_2$ in the peak effect (PE) region at 4.5\,K. The inset in Fig.~11(b) shows the complete two quadrant isothermal $M$-$H$ loop at the same temperature. Note that there is a current decay across the PE region, and the filled square data points show the $M$ values after an interval of 1000\,s. The dotted line is a guide to eye for a section of the decayed PE loop. The inset shows a typical $M(t)$ profile at 20\,kOe. In Fig.~11(b), we show a plot of $S(H)$ across the PE region in CeRu$_2$. There is a non-monotonic modulation in $S$ across the PE. Following observations are worth noting: (i) The $M$-$H$ loop is reversible (no hysteresis) over a considerable field range (5\,kOe to 19\,kOe at 4.5\,K) prior to the onset of PE ($H_{\rm PE}^{\rm on}$) in CeRu$_2$ (see inset in Fig.~11(b)). In this region, $J_c$ is very small ($<100 A/cm^2$) with no measurable time evolution. Consequently, $S$ is essentially zero in this region (see Fig.~11(b)). In contrast, $J_c$ is non-zero prior to the onset of SMP ($H_{\rm SMP}^{\rm on}$) in FeSe$_{0.5}$Te$_{0.5}$. There is a significant time decay response and consequently, the baseline value of $S$ is non-zero prior to $H_{\rm SMP}^{\rm on}$, (ii) $S$ varies non-monotonically across both the SMP in FeSe$_{0.5}$Te$_{0.5}$ as well as across the PE in CeRu$_2$, (iii) The field value for the maximum in $S$ does not exactly coincide with $H_{\rm SMP}^{\rm p}$ (or, $H_{\rm PE}^{\rm p}$). However, it is worthwhile to note that the ballpark field region (between $H_{\rm PE}^{\rm on}$ and $H_{\rm PE}^{\rm p}$) where, the value of $S$ peaks in the case of CeRu$_2$, coincides with the field value at which a jump in equilibrium magnetization (implying a first order transition) across PE in CeRu$_2$ was observed by Tulapurkar {\it et al} \cite{ashwin}.
The PE in CeRu$_2$ is associated with an order-disorder transition in vortex matter. The occurence of SMP in weakly pinned type-II superconductors has also been conjured as a manifestation of a disorder driven order-disorder transformation from collective pinned elastic state to small bundle pinning plastic region \cite{proz1}. In a typical field ramp experiment, vortices are injected into the sample through inhomogeneous surface barriers leading to the formation of transient disordered vortex states (TDVS) within the sample \cite{kalisky, adtphysica}. Here, it is interesting to understand the non-monotonic modulation in $S$ in FeSe$_{0.5}$Te$_{0.5}$ within the picture of annealing of these TDVS across the order-disorder phase transition in vortex lattices. Kalisky {\it et al} \cite{kalisky} reported the role of TDVS on the magnetic relaxation across the SMP in single crystals of Bi2212. The non-monotonicity of $S$ with $H$ across the SMP is evident in their observations. The observation of a similar non-monotonic variation of S across SMP in our single crystal of FeSe$_{0.5}$Te$_{0.5}$ fortifies the plausibility of the occurence of a similar order-disorder transformation. Closeness in the numerical values of $S$ across the SMP in FeSe$_{0.5}$Te$_{0.5}$ and that across PE in CeRu$_2$ could also point to similarity in the nature of current decay across SMP and PE features. However, a more conclusive inference can not be arrived at from our present data.

To summarize, we have explored the phenomenon of SMP across ($H$,$T$) vortex phase diagram in FeSe$_{0.40}$Te$_{0.60}$. There are considerable differences in the observed scenario in the cases for fields parallel and perpendicular to the $c$-axis, respectively. This in turn suggests the possible role of crystalline anisotropy and its effect on the vortex lattice structures. Attempt is made to quantify the observations within the Dew-Hughes scenario \cite{dew1, dew2}. Such an analysis appears to suggest that point pins alone can not rationalize the observed field variation of the pinning force density. A comparison of the observed $J_c(t)$ values is made with the anticipated temperature variations within the $\delta l$ and $\delta T_{\rm c}$ pinning scenarios. We find strong evidence for $\delta l$ pinning leading to small bundle vortex pinning by randomly distributed weak pinning centers. Experimental signatures via magnetic relaxation studies suggesting the possibility of an order-disorder transformation across SMP is seen in the case of FeSe$_{0.5}$Te$_{0.5}$ and a comparison with the case of PE phenomenon in CeRu$_2$ is drawn. The observations present a strong case for a dedicated local probe measurement (either via magneto-optical imaging or, by a suitable magnetic decoration technique) which is expected to shed more light on the finer details of the pinning mechanism in Fe-based superconductors.

We would like to thank Prof. Y. Onuki for providing the single crystal of CeRu$_2$. GB, MRL and CVT wishes to acknowledge financial support for this research from EPSRC, UK. CVT would like to acknowledge the Department of Science and Technology for partial support through the project IR/S2/PU-10/2006. AKY would like to thank CSIR, India for SRF grant. ADT acknowledges the Indian Institute of Technology, Bombay for financial support in the form of Institute Post-Doctoral Fellowship. 

\newpage
\begin{figure}
\includegraphics[scale=0.5,angle=0]{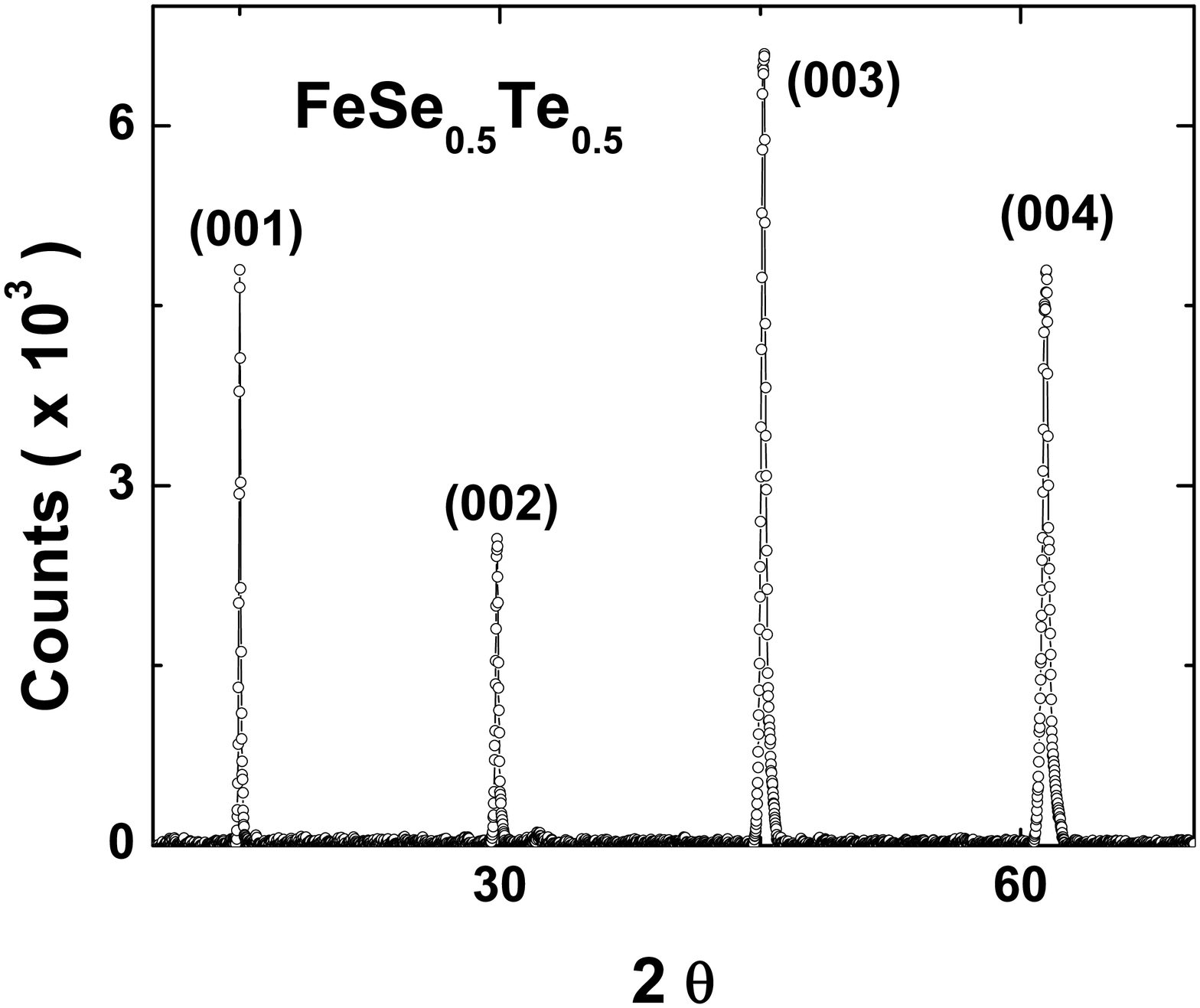}
\caption{The x-ray diffraction pattern with identification of $(00l)$ lines for a cleaved piece of single crystal of FeSe$_{0.5}$Te$_{0.5}$ (see text for details).}
\end{figure}
\begin{figure}
\includegraphics[scale=0.5,angle=0]{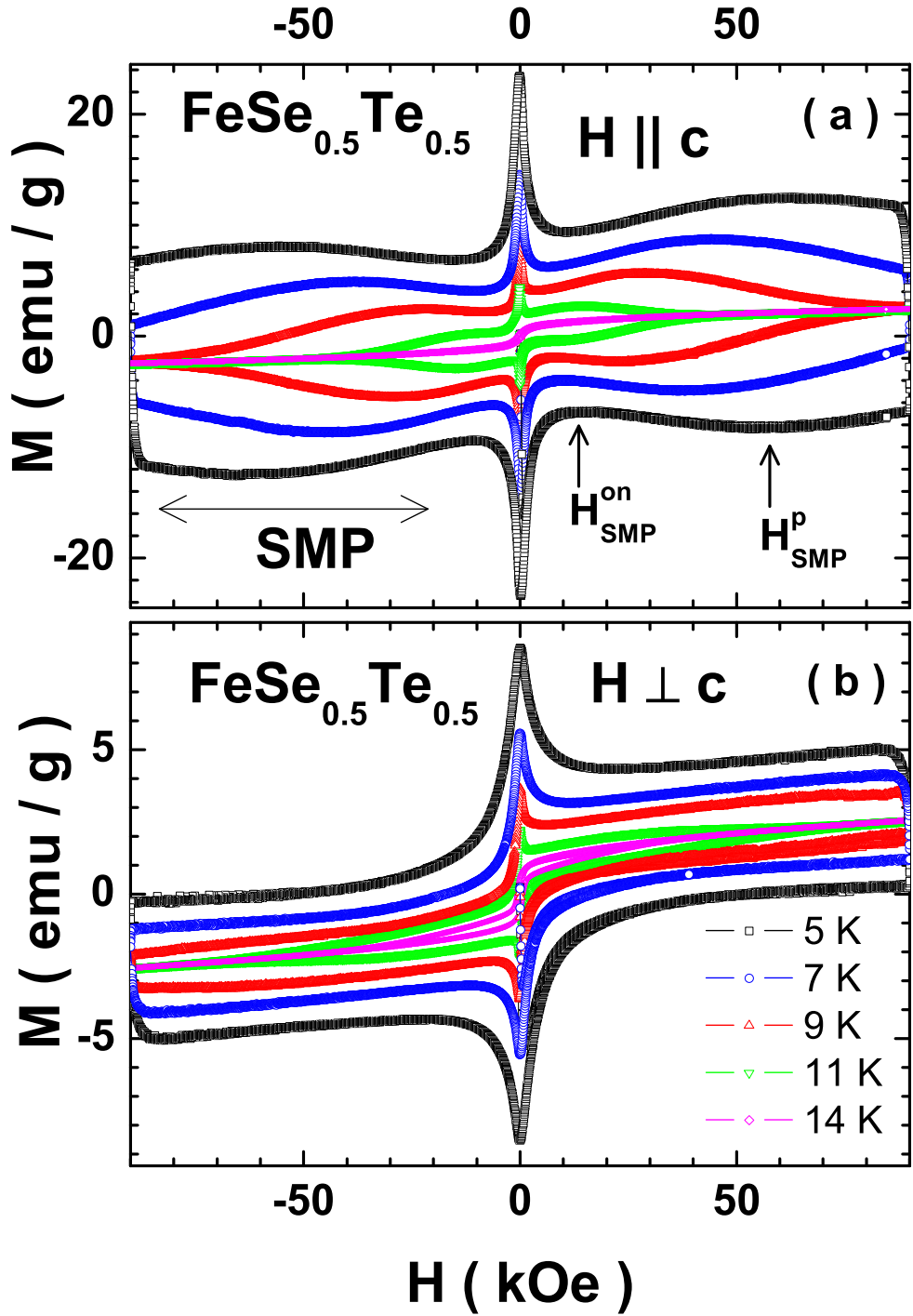}
\caption{(Color online) Isothermal $M$-$H$ measurements in a single crystal of FeSe$_{0.5}$Te$_{0.5}$ at various temperatures as indicated for the case of: (a) $H\parallel {\rm c}$, and (b) $H\perp {\rm c}$.}
\end{figure}
\begin{figure}
\includegraphics[scale=0.6,angle=0]{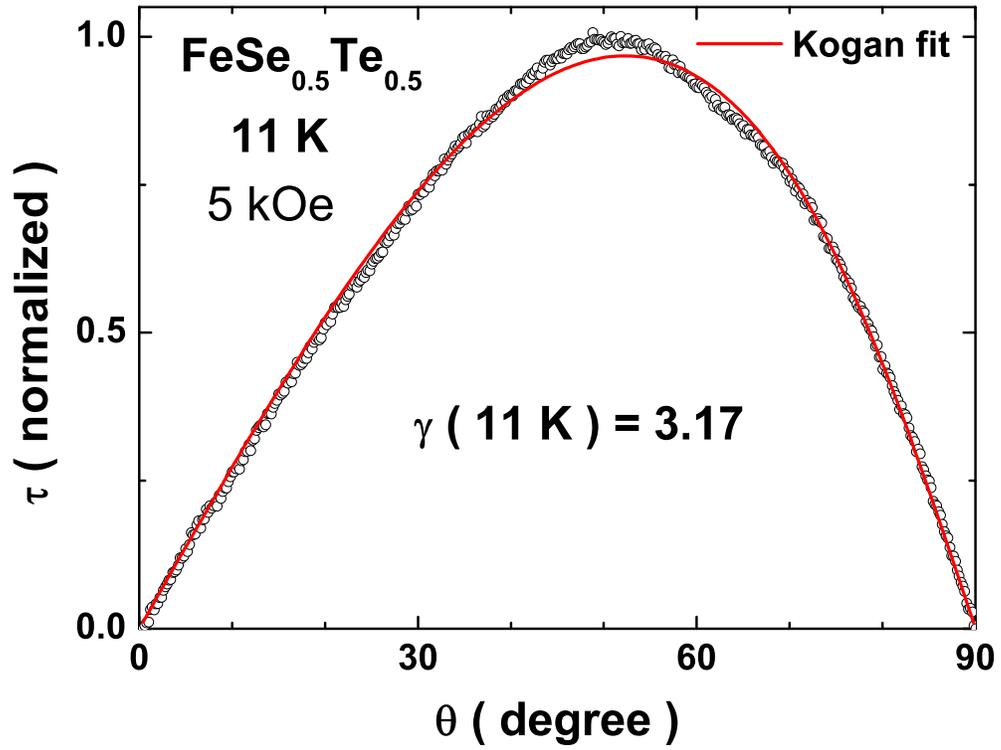}
\caption{(Color online) Torque magnetometry data (see text) as a function of $\theta$ in a single crystal of FeSe$_{0.5}$Te$_{0.5}$ at 11\,K and 5\,kOe. The solid line shows the fit to the Kogan equation \cite{kogan88, farrel}. }
\end{figure}
\begin{figure}
\includegraphics[scale=0.6,angle=0]{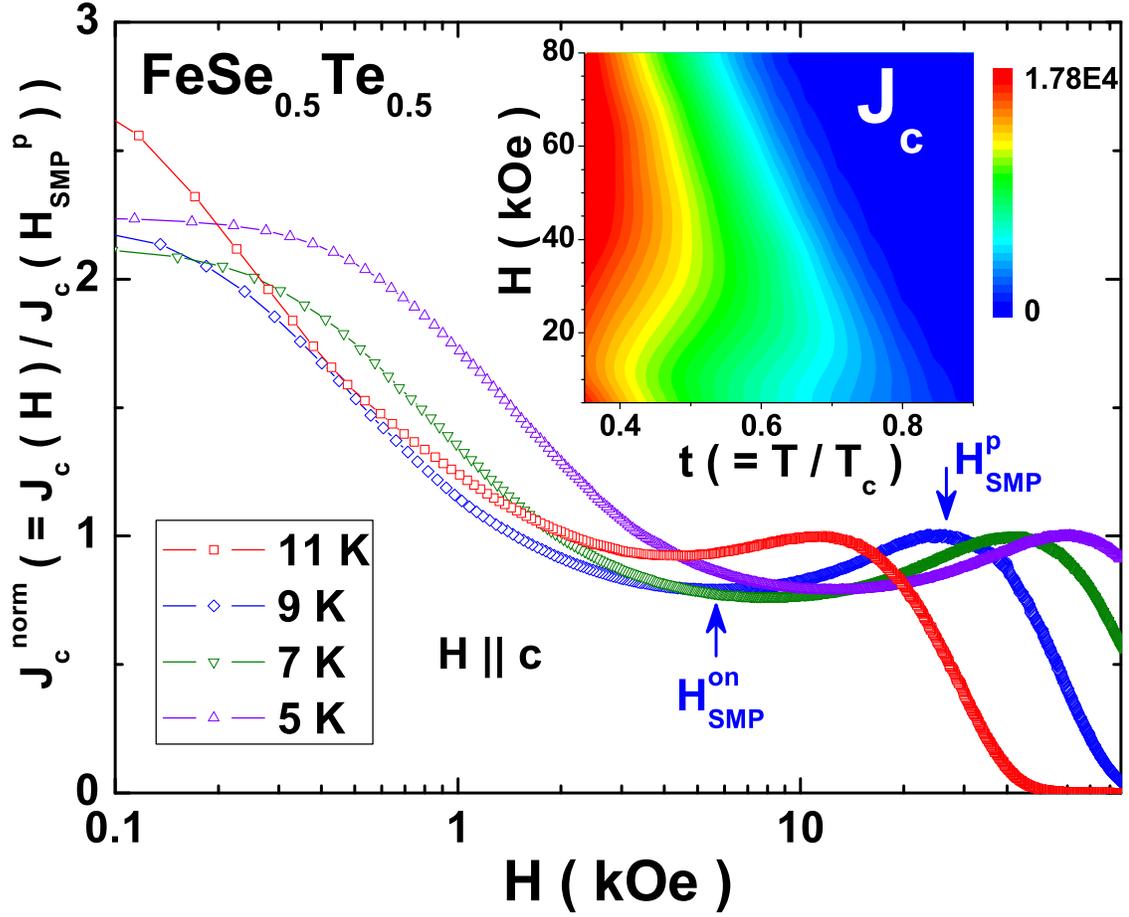}
\caption{(Color online) $J_c^{\rm norm}(H)$ ($=J_c(H) / J_c(H_{\rm SMP}^{\rm p}$) in a single crystal of FeSe$_{0.5}$Te$_{0.5}$ at different temperatures, as indicated. Inset shows a color scale plot of the critical current density, $J_c(H,t)$, where t ($=T/T_{\rm c}$) is the reduced temperature. The characteristics of SMP can be clearly seen for reduced temperature values in the interval $0.35 < t < 0.50$.}
\end{figure}
\begin{figure}
\includegraphics[scale=0.6,angle=0]{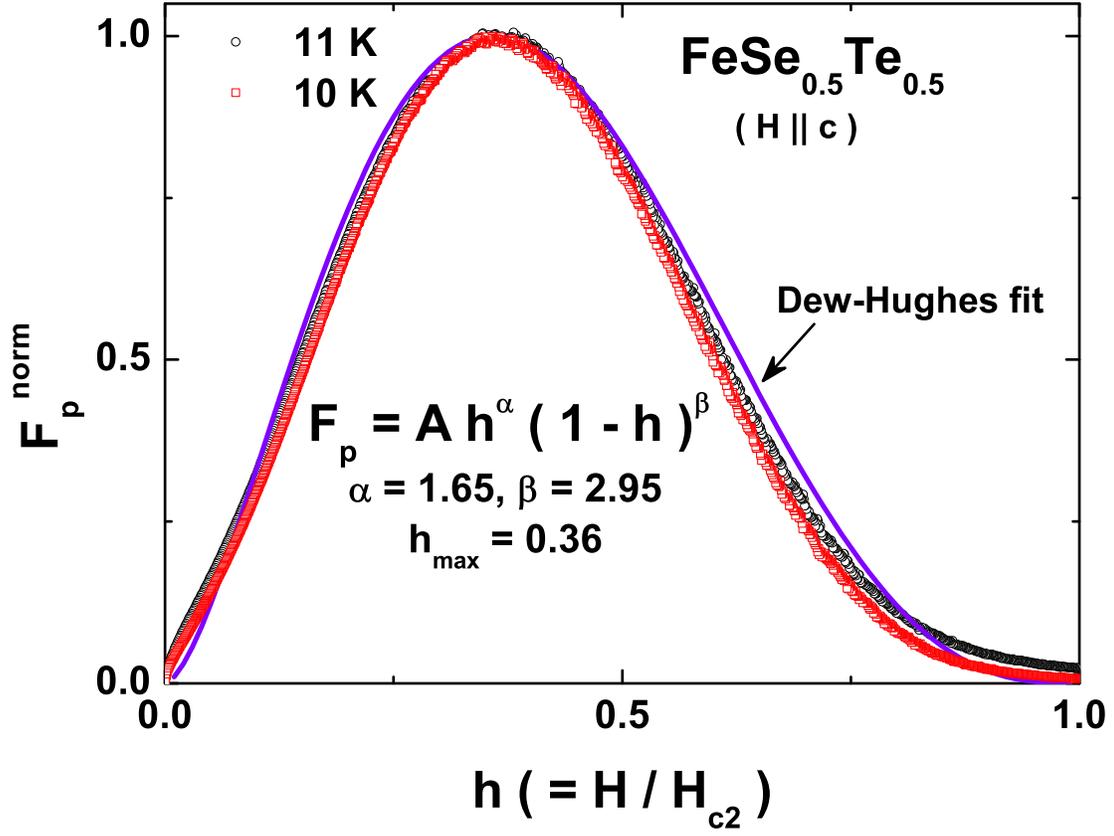}
\caption{(Color online) Normalized pinning force density, $F_{\rm p}^{\rm norm}$ ($=F_{\rm p} (H) / F_{\rm p}^{\rm max} (H)$ as a function of reduced field, $h$ ($=H/H_{c2}$). The solid line shows the fit to the Dew-Hughes's formula $F_{\rm p} \sim h^{\alpha}(1-h)^{\beta}$ \cite{dew1, dew2}.}
\end{figure}
\begin{figure}
\includegraphics[scale=0.4,angle=0]{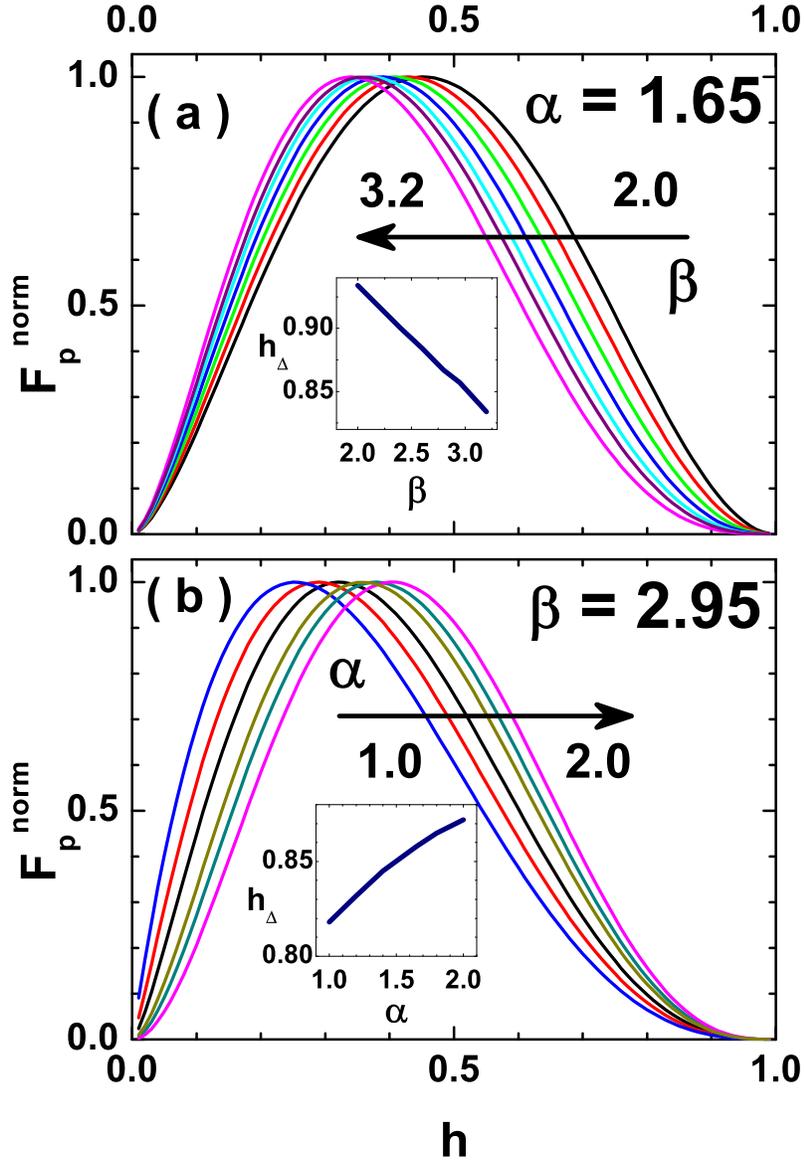}
\caption{(Color online) Variation of $F_{\rm p}^{\rm norm}$ as a function of reduced field, $h$ ($=H/H_{c2}$) within the Dew-Hughes formalism for the cases: (a) $\alpha$ fixed to the best fit value and $\beta$ varied, and (ii) $\beta$ fixed to the best fit value and $\alpha$ varied. The insets in panels (a) and (b) shows the variation of $h_{\rm \Delta}$ as the magnitudes of the exponents $\alpha$ and $\beta$ are varied about the best fit values. Here $h_{\rm \Delta}$ is the reduced field value at which $F_{\rm p}^{\rm norm} (h_{\rm \Delta})$ attains a value of 0.05 (see text for details).}
\end{figure}
\begin{figure}
\includegraphics[scale=0.5,angle=0]{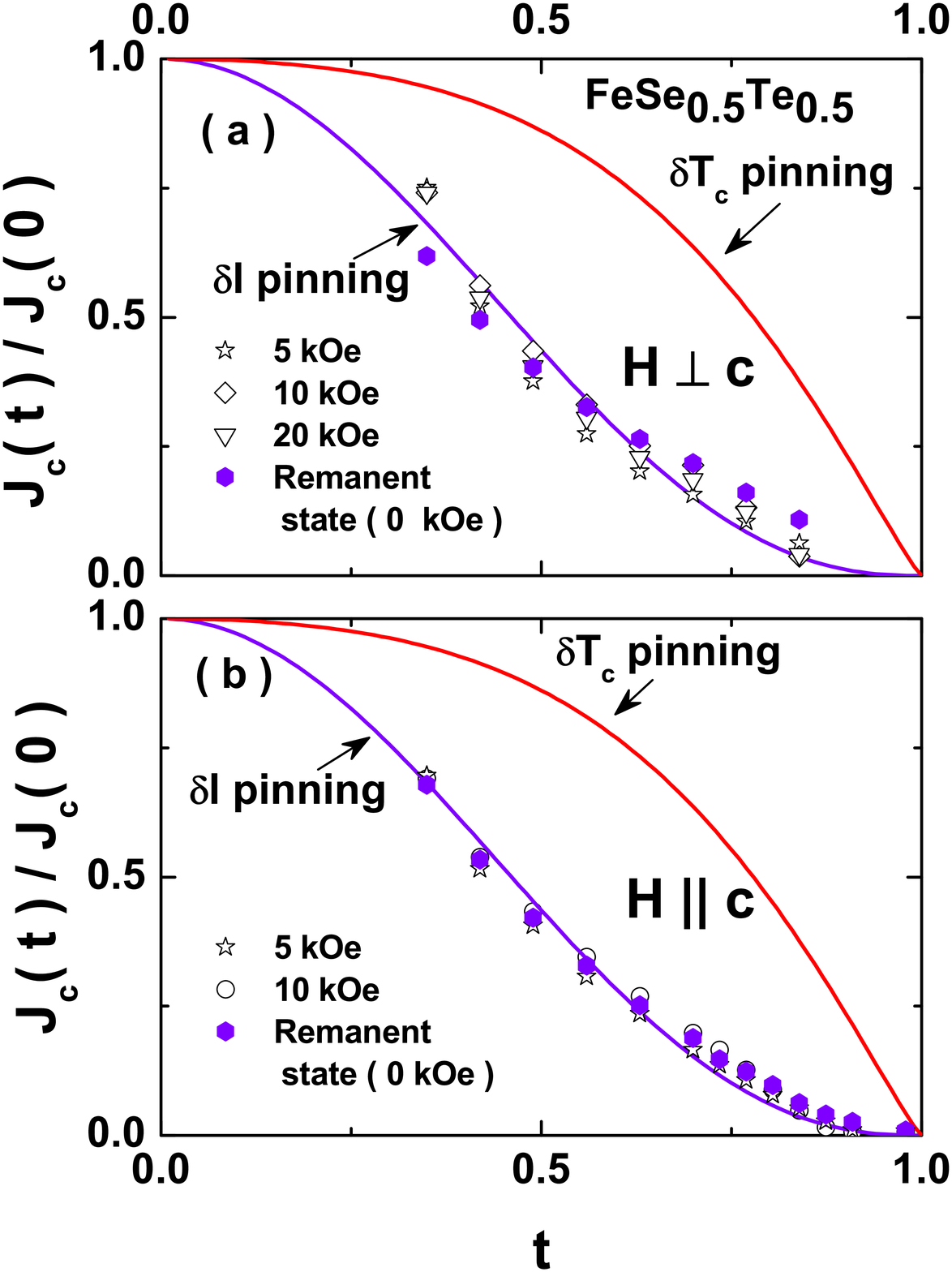}
\caption{(Color online) $J_{\rm c} (t)/J_{\rm c} (0)$ data at: (a) 0\,kOe (filled hexagons), 5\,kOe (open stars) and 10\,kOe (open circles) for $H\perp {\rm c}$, and (b) 0\,kOe (filled hexagons), 5\,kOe (open stars), 10\,kOe (open diamonds) and 20\,kOe (open triangles) for $H\parallel {\rm c}$ in a single crystal of FeSe$_{0.5}$Te$_{0.5}$. The theoretical estimates for $\delta T_{\rm c}$ and $\delta l$ pinning are shown by bold lines.}
\end{figure}
\begin{figure}
\includegraphics[scale=0.4,angle=0]{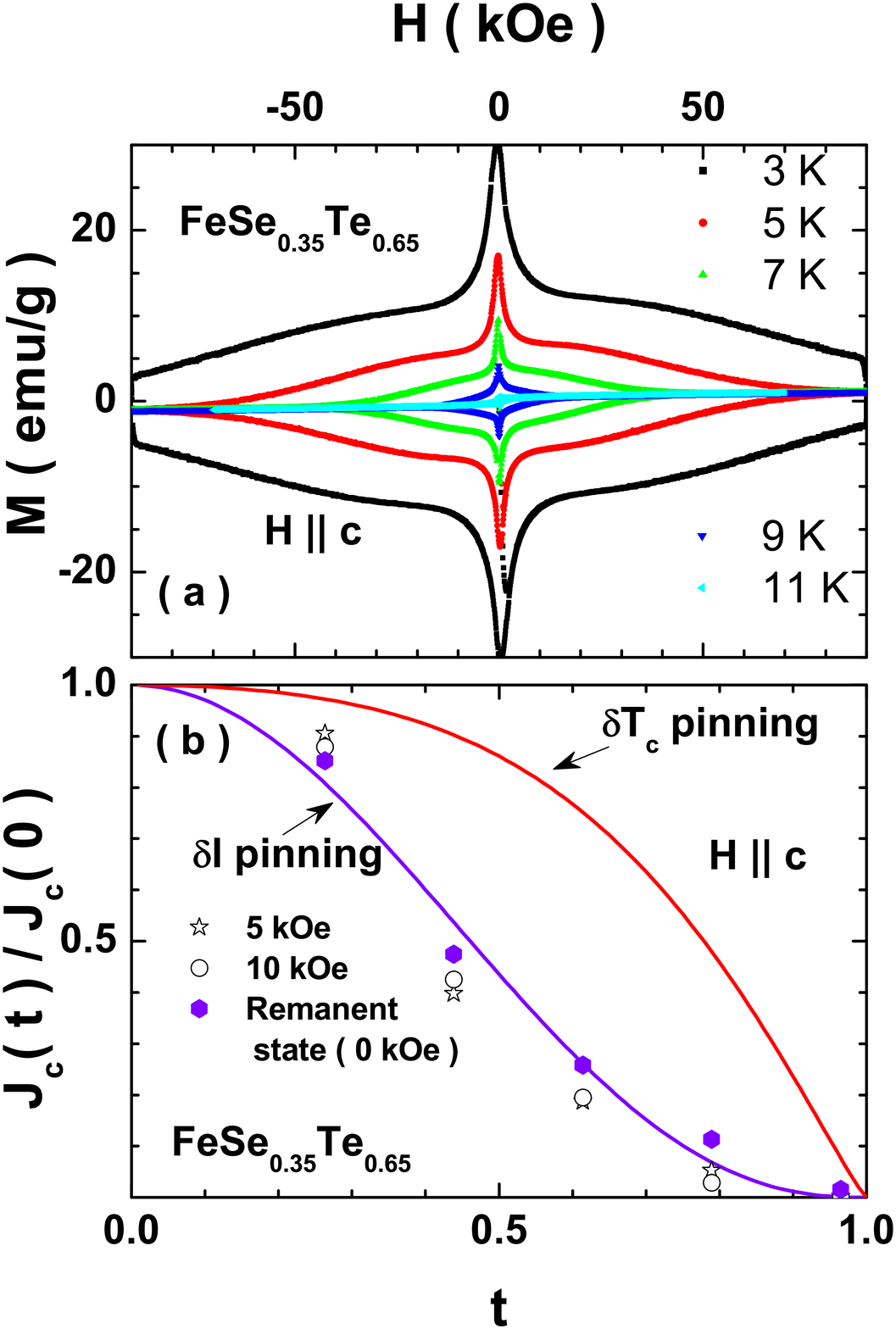}
\caption{(Color online) (a) Isothermal $M$-$H$ measurements in a single crystal of FeSe$_{0.35}$Te$_{0.65}$ at various temperatures as indicated for $H\parallel {\rm c}$, (b) Normalized $J_{\rm c} (t)$ data at 0\,kOe (filled hexagons), 5\,kOe (open stars) and 10\,kOe (open circles) in a single crystal of FeSe$_{0.35}$Te$_{0.65}$. The theoretical estimates for $\delta T_{\rm c}$ and $\delta l$ pinning are shown by bold lines.}
\end{figure}
\begin{figure}
\includegraphics[scale=0.6,angle=0]{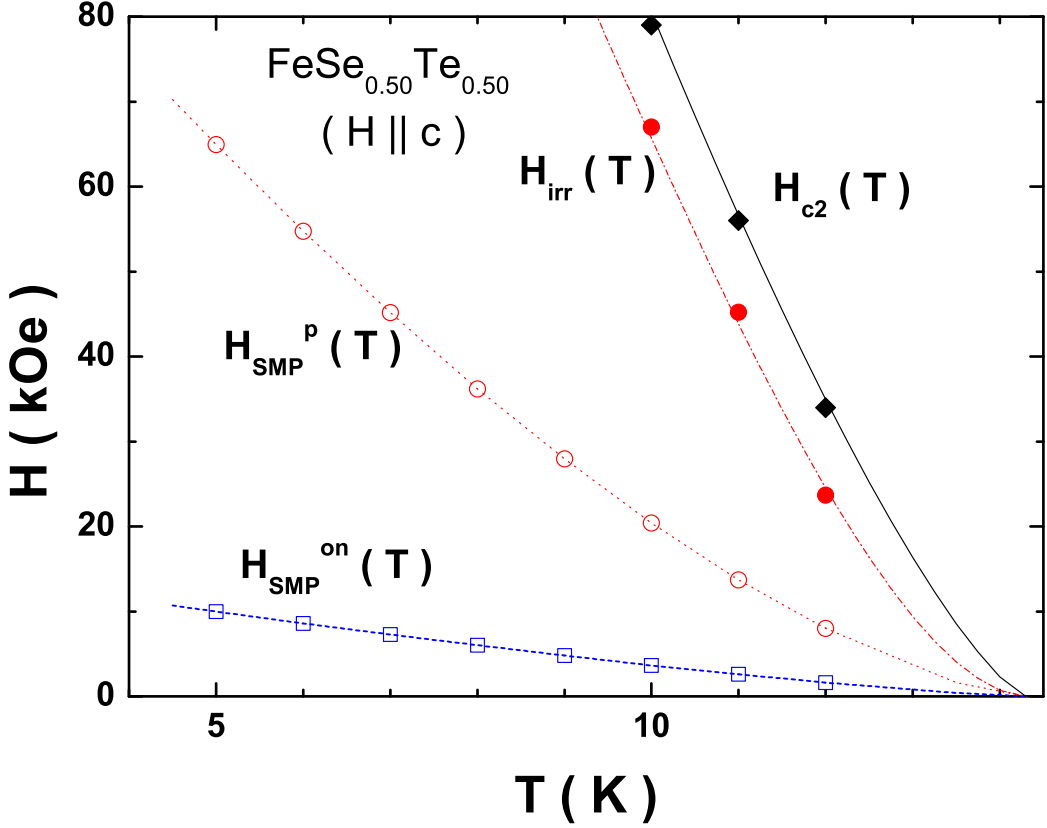}
\caption{(Color online) Plot of the temperature variations of different characteristic fields identified in the $M$-$H$ scans in the form of a magnetic phase diagram for FeSe$_{0.5}$Te$_{0.5}$ ($H \parallel c$). Least square fits to the expression $H_x(T) = H_x(0)[1-(\frac{T}{T_{\rm c}})^p]^n$ are also shown for these characteristic fields. The least square fits yield the following results for the various characteristic fields : (i) for $H_{\rm SMP}^{\rm on}(T)$, $p=1$, $n=\frac{4}{3}$ and $H_{\rm SMP}^{\rm on}(0) = 17.45~kOe$, (ii) for $H_{\rm SMP}^{\rm p}(T)$, $p=1$, $n=\frac{3}{2}$ and $H_{\rm SMP}^{\rm p}(0) = 123.8~kOe$, (iii) for $H_{\rm irr}(T)$, $p=2$, $n=\frac{9}{5}$ and $H_{\rm irr}(0) = 220~kOe$, and (iv) for $H_{\rm SMP}^{\rm on}(T)$, $p=1$, $n=\frac{4}{3}$ and $H_{\rm SMP}^{\rm on}(0) = 400~kOe$.}
\end{figure}
\begin{figure}
\includegraphics[scale=0.5,angle=0]{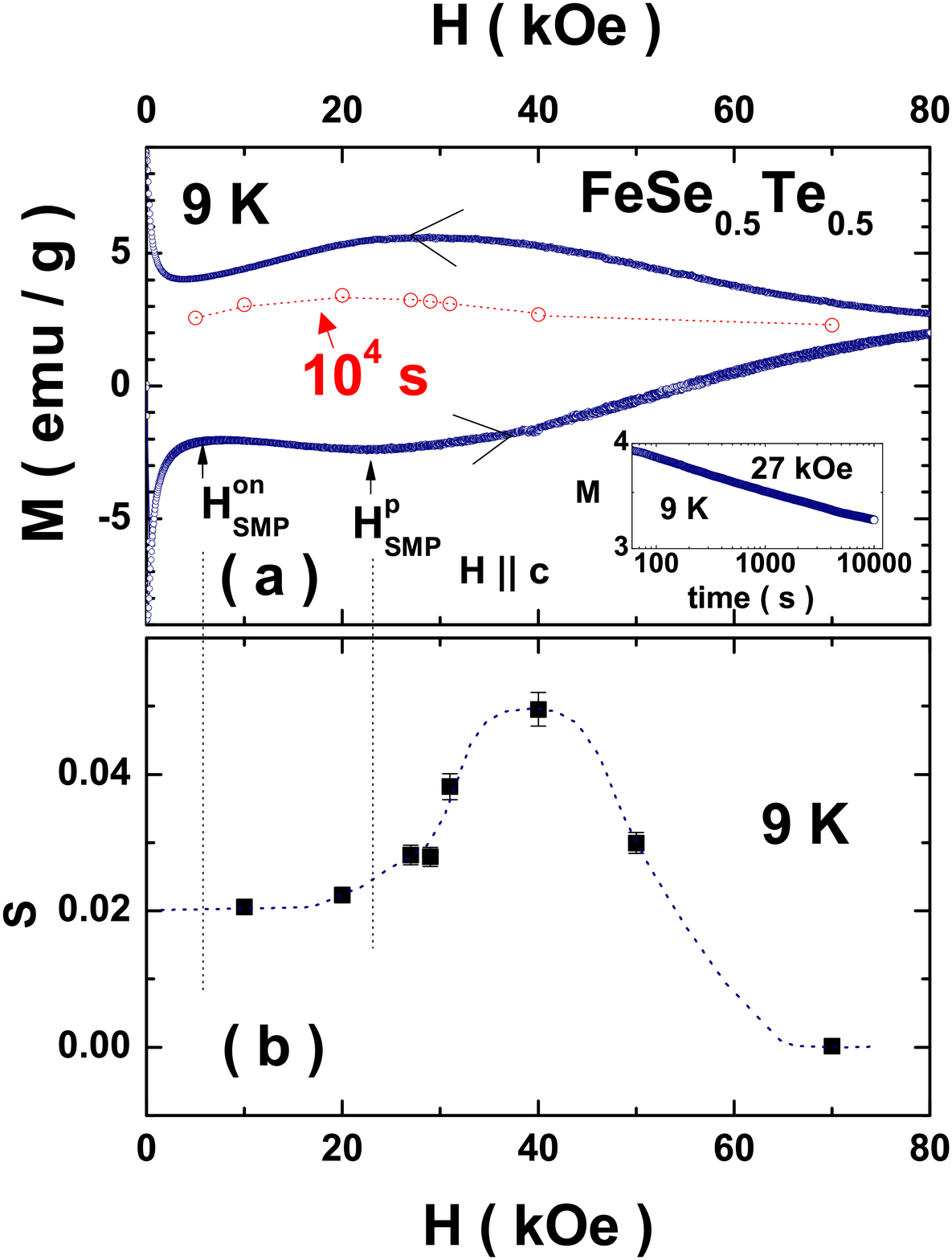}
\caption{(Color online) Relaxation across the fishtail effect in FeSe$_{0.5}$Te$_{0.5}$ : (a) portion of isothermal $M$-$H$ loop at 9\,K ($t=T/T_{\rm c}=0.63$). The inset in panel (a) shows the time decay ($M(t)$) data measured up to 10$^4$\,s at 27\,kOe and 9\,K. The open red symbols show the magnetization values at 10$^4$\,s at respective fields with the dotted line being a guide to eye for the relaxed return leg of the $M$-$H$ loop at 10$^4$\,s, (b) $S$ parameter as a function of field at 9\,K.}
\end{figure}
\begin{figure}
\includegraphics[scale=0.5,angle=0]{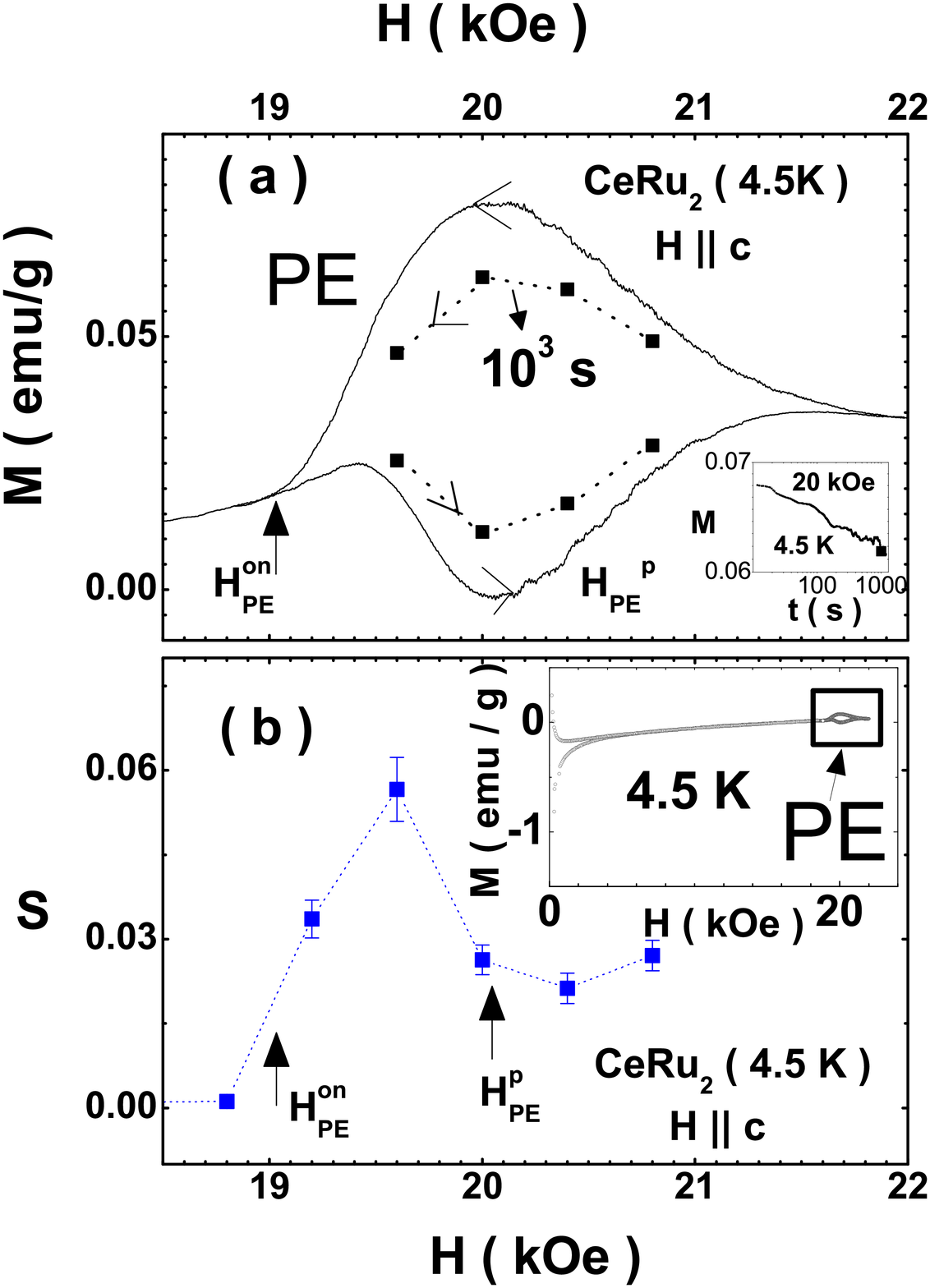}
\caption{(Color online) Relaxation data across the Peak Effect (PE) in CeRu$_2$ : (a) portion of isothermal $M$-$H$ loop at 4.5\,K in the PE region. The inset in panel (a) shows the time decay ($M(t)$) data measured up to 10$^3$\,s at 20\,kOe and 4.5\,K, (b) $S$ parameter as a function of field at 4.5\,K. The inset in panel (b) shows the 2 quadrant isothermal $M$-$H$ loop obtained at 4.5\,K with the PE being highlighted by the boxed region.}
\end{figure}
\end{document}